\documentclass{moriond}

\usepackage{amsmath,amsfonts}
\usepackage{xspace}
\usepackage{multirow}
\usepackage[square,sort,comma,numbers]{natbib}
\def\Journal#1#2#3#4{{#1} {\bf #2}, #3 (#4)}

\def\NIMA{{\em Nucl. Instrum. Methods} A}

\def\PRL{\em Phys. Rev. Lett.}
\def\PRD{{\em Phys. Rev.} D}

\def\be{\begin{equation}}
\def\ee{\end{equation}}
\def\bea{\begin{eqnarray}}
\def\eea{\end{eqnarray}}

\def\nue        {\ensuremath{\nu_e}\xspace}

\def\num        {\ensuremath{\nu_\mu}\xspace}

\newcommand{\tab}[1]     {Table~\ref{#1}}

\newcommand{\Photo}{}

\begin{document}
\vspace*{4cm}
\title{Oscillation results from T2K}

\author{ Patrick de Perio for the T2K Collaboration}

\address{Department of Physics, University of Toronto, 60 St. George Street,
Toronto M5S 1A7, Canada}

\maketitle

\abstracts{
The T2K collaboration has combined the \num disappearance and \nue
appearance data in a three-flavor neutrino oscillation analysis. A
Markov chain Monte Carlo (MCMC) results in estimates of the
oscillation parameters and 1D 68\% 
Bayesian credible intervals (CI) at $\delta_{CP} = 0$ as follows: 
$\sin^{2}\theta_{23} = 0.520^{+0.045}_{-0.050}$,
$\sin^{2}\theta_{13} = 0.0454^{+ 0.011}_{-0.014}$ and $|\Delta
m^{2}_{32}| = 2.57\pm0.11$, with the point of highest posterior
probability in the inverted hierarchy. Recent
measurements of $\theta_{13}$ from reactor neutrino experiments are
combined with the T2K data resulting in the following estimates: 
$\sin^{2}\theta_{23} = 0.528^{+0.055}_{-0.038}$,
$\sin^{2}\theta_{13} = 0.0250 \pm 0.0026$ and $|\Delta
m^{2}_{32}| = 2.51\pm0.11$, with the point of highest posterior
probability in the normal hierarchy. Furthermore, the data exclude
values of $\delta_{CP}$ between 0.14$\pi$--0.87$\pi$ with 90\%
probability.  
}

\section{Neutrino Oscillations}

The standard three neutrino model relates the flavor states of
neutrinos to the mass states through the PMNS mixing
matrix~\cite{bib:mns,bib:ponte}.  
This matrix is parameterized by three mixing angles ($\theta_{12}$,
$\theta_{13}$, $\theta_{23}$) and one CP violating phase
($\delta_{CP}$). The flavor transition probabilities depend on these
parameters as well as mass-squared splittings ($\Delta m^2_{ij}$) and
the mass hierarchy (MH), where $\Delta m^2_{31} > 0$ is defined as the
normal hierarchy (NH) and $\Delta m^2_{31} < 0$ the inverted hierarchy
(IH). All the parameters except $\delta_{CP}$, the $\theta_{23}$
octant (maximal or non-maximal) and the MH have been previously  
measured as summarized in Reference~\cite{bib:pdg2013}.

The T2K experiment~\cite{bib:t2k}
aims to measure $\theta_{23}$ and $|\Delta m^2_{31}| \approx
|\Delta m^2_{32}|$ precisely via \num
disappearance, with the survival probability given by
\begin{equation}
\label{eq:numu_disapp}
  P_{{\nu}_{\mu} \rightarrow {\nu}_{\mu}} \approx 
  1 - (\cos^4\theta_{13} \sin^22\theta_{23} +
  \sin^22\theta_{13}\sin^2\theta_{23})\sin^2\mathit{\Delta},
\end{equation}
where $\mathit{\Delta} = {\Delta m^2_{31}L}/{(4E)}$, and $L$ and $E$ are
the neutrino flight length and energy, respectively. T2K also
measures \num to \nue appearance with the probability expanded in
$\alpha = {\Delta m^2_{21}}/{\Delta m^2_{31}}$ as~\cite{bib:freund_probmatter} 
\begin{equation}
\label{eq:3flav_app_prob}
P_{{\nu}_{\mu} \rightarrow {\nu}_{e}} \approx 
\sin^2 2\theta_{13}T_1 - \alpha\sin2\theta_{13}T_2 + \alpha\sin2\theta_{13}T_3 + \mathcal{O}(\alpha^2),
\end{equation}where\begin{eqnarray}
&T_1 = \sin^2\theta_{23}\frac{\sin^2[(1-x)\mathit{\Delta}]}{(1-x)^2}, \\
&T_2 = \sin\delta \sin2\theta_{12}\sin2\theta_{23}\sin\mathit{\Delta}\frac{\sin(x\mathit{\Delta})}{x}\frac{\sin[(1-x)\mathit{\Delta}]}{(1-x)},\\
&T_3 = \cos\delta \sin2\theta_{12}\sin2\theta_{23}\cos\mathit{\Delta}\frac{\sin(x\mathit{\Delta})}{x}\frac{\sin[(1-x)\mathit{\Delta}]}{(1-x)},
\end{eqnarray}
and $x$ is the correction due to the matter
effect, which is positive (negative) for the NH (IH). 
Thus, T2K is sensitive to $\theta_{13}$ and can explore
$\delta_{CP}$, especially through the CP-odd term, $T_2$.

Since there are common parameters in
Equations~\ref{eq:numu_disapp}~and~\ref{eq:3flav_app_prob}, a combined
analysis of the \num disappearance and \nue appearance samples rather
than two separate analyses~\cite{bib:t2k_nueapprun1-4,bib:t2k_numurun1-4} can improve the 
oscillation parameter measurements, in principle.


\section{The T2K Experiment}
\label{sec:t2k}

An intense and high purity \num beam is produced at J-PARC by colliding 
a 30~GeV proton beam with a graphite target, then focusing the
resulting charged hadrons by magnetic horns prior to decay into
neutrinos. The far detector, Super-Kamiokande (SK), is situated
2.5$^\circ$ off-axis from the neutrino beam resulting in a narrow-band
energy spectrum peaked at 0.6~GeV, which maximizes the \nue appearance
probability at a baseline of $L = 295$~km and minimizes high energy
backgrounds. This baseline corresponds to a matter effect correction
of $|x| \approx 5\%$.
The near detector complex, 280~m from the average neutrino production
point, consists of an on-axis (INGRID) and off-axis (ND280) detector
to constrain the beam direction, and neutrino flux and cross sections
(xsec.), respectively.  

The flux prediction is based on simulations tuned and constrained by
hadron production data from the NA61/SHINE experiment and in-situ
proton beam monitoring. The NEUT simulation package is used for the
neutrino interaction model, with prior constraints based on external
neutrino, pion and nucleon scattering cross section measurements. 

The SK analysis uses a single-ring sample, which enhances charged
current (CC) quasi-elastic (QE) events, separated into $\mu$-like (\num)
and $e$-like (\nue) sub-samples. The ND280 analysis selects charged current
(CC) \num interactions and separates the sample based on the number of
reconstructed pions and decay electrons: CC0$\pi$, CC1$\pi$ and CC
Other. These topologies provide a strong constraint on the flux and 
interaction model governing CCQE scattering and resonant pion
production, the signal and main background to the SK analysis,
respectively. The reduction in the uncertainty on the SK predicted
event rates due to the ND280 data is shown in
Table~\ref{tab:nsk_syst_summary}. The SK and ND280 detector errors are
constrained by calibration data and control samples such as cosmic
rays and atmospheric neutrinos. More details of the SK and ND280
analyses can be found in previous T2K
publications e.g.~\cite{bib:t2k_nueapprun1-4,bib:t2k_numurun1-4} and the 
references therein. 

\begin{table}[htbp]
\caption[Summary of Systematic Errors on SK Event Rate]{
  Summary of the effect of the systematic errors on the SK \nue and \num
  candidate total event rates. The prior uncertainties in () brackets
  do not include the ND280 data.
}
\vspace{0.2cm}
\begin{center}
\begin{tabular}{|c|cc|}
\hline
\multirow{2}{*}{Systematic Error Source} & \multicolumn{2}{c|}{Relative Uncertainty (\%)}\\
   & \num Candidates  & \nue Candidates   \\
\hline
Flux \& Xsec. ND280 Constrained (Prior) & 2.7 (21.7) & 3.1 (26.0) \\
Xsec. ND280 Independent                         & 5.0        & 4.7        \\
Pion Hadronic Interactions                      & 3.5        & 2.3        \\
SK Detector                                     & 3.6        & 2.9        \\
\hline                                                                    
Total ND280 Constrained (Prior)         & 7.6 (23.4) & 6.8 (26.8) \\
\hline
\end{tabular}
\label{tab:nsk_syst_summary}
\end{center}
\end{table}

\section{Oscillation Analysis}

\begin{figure}[htbp] 
   \centering
   \includegraphics[width=\textwidth]{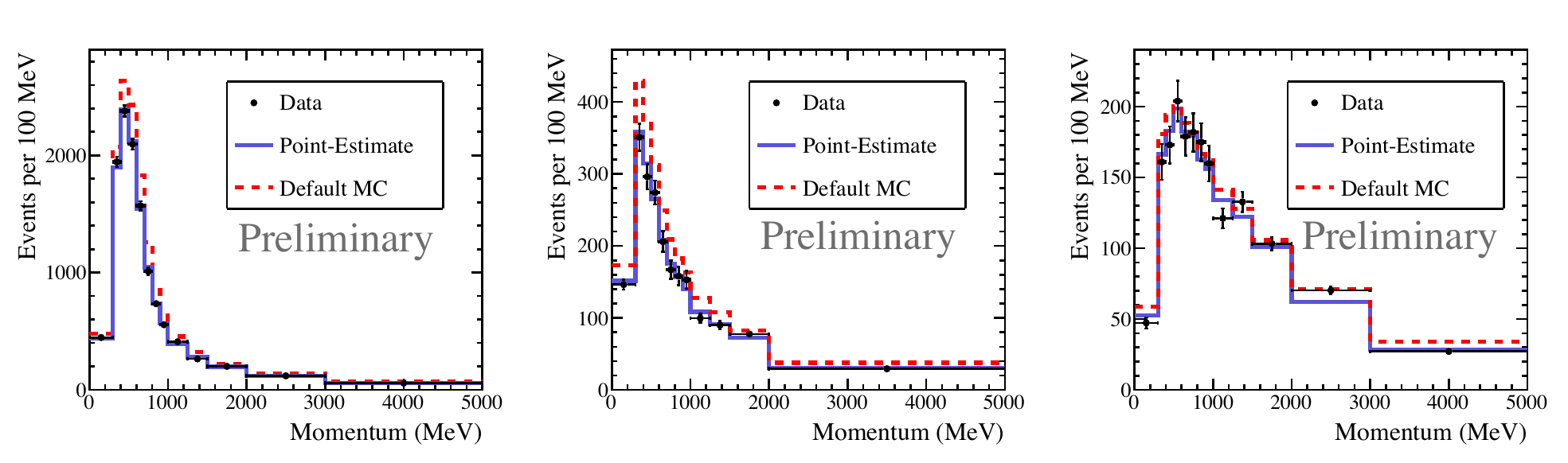} 
\vspace{-0.4in}
   \caption[Data and Point-Estimate MC Event Rates for ND280 Samples in
   $p_\mu$]{
     The ND280 CC0$\pi$ (left), CC1$\pi$ (middle) and CC Other (right)
     event rates projected onto muon momentum for data, the default MC
     and point-estimates. 
   }
   \label{fig:pfmom}

\par\vspace{\intextsep}

   \centering
   \includegraphics[width=\textwidth]{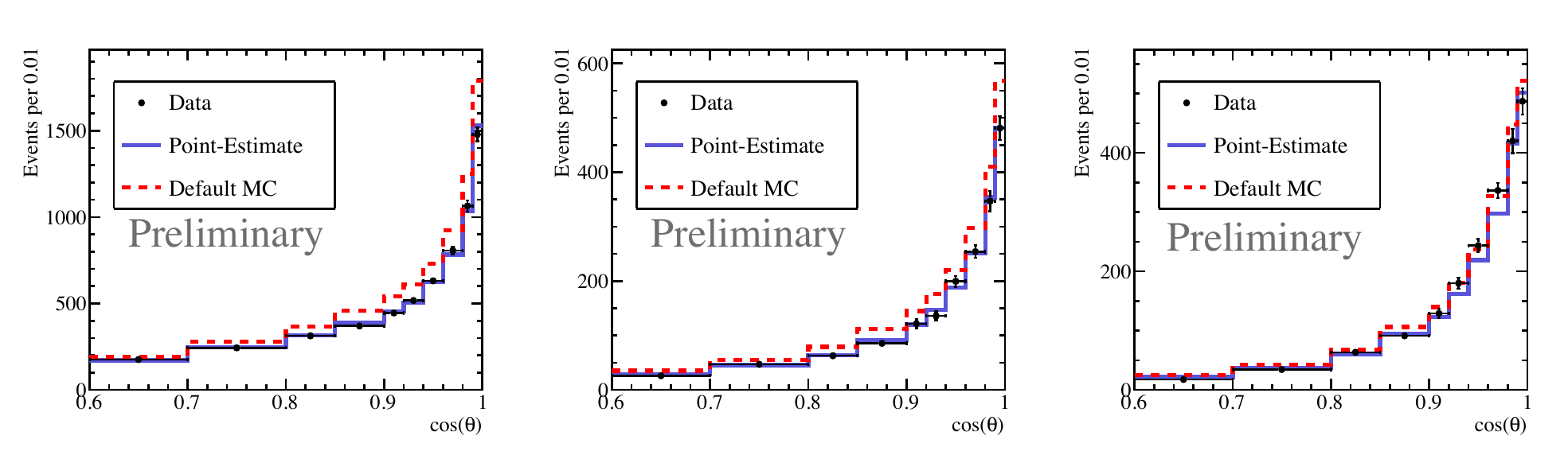} 
\vspace{-0.45in}
   \caption[Data and Point-Estimate Event Rates for ND280 Samples in $\cos\theta$]{
     The ND280 CC0$\pi$ (left), CC1$\pi$ (middle) and CC Other (right)
     event rates projected onto $\cos\theta$ (muon angle relative to
     the beam) for data, the default MC and point-estimates.
} 
\label{fig:pfang}
\end{figure}

\begin{figure}[htbp]
\centering
\includegraphics[width=4.8in]{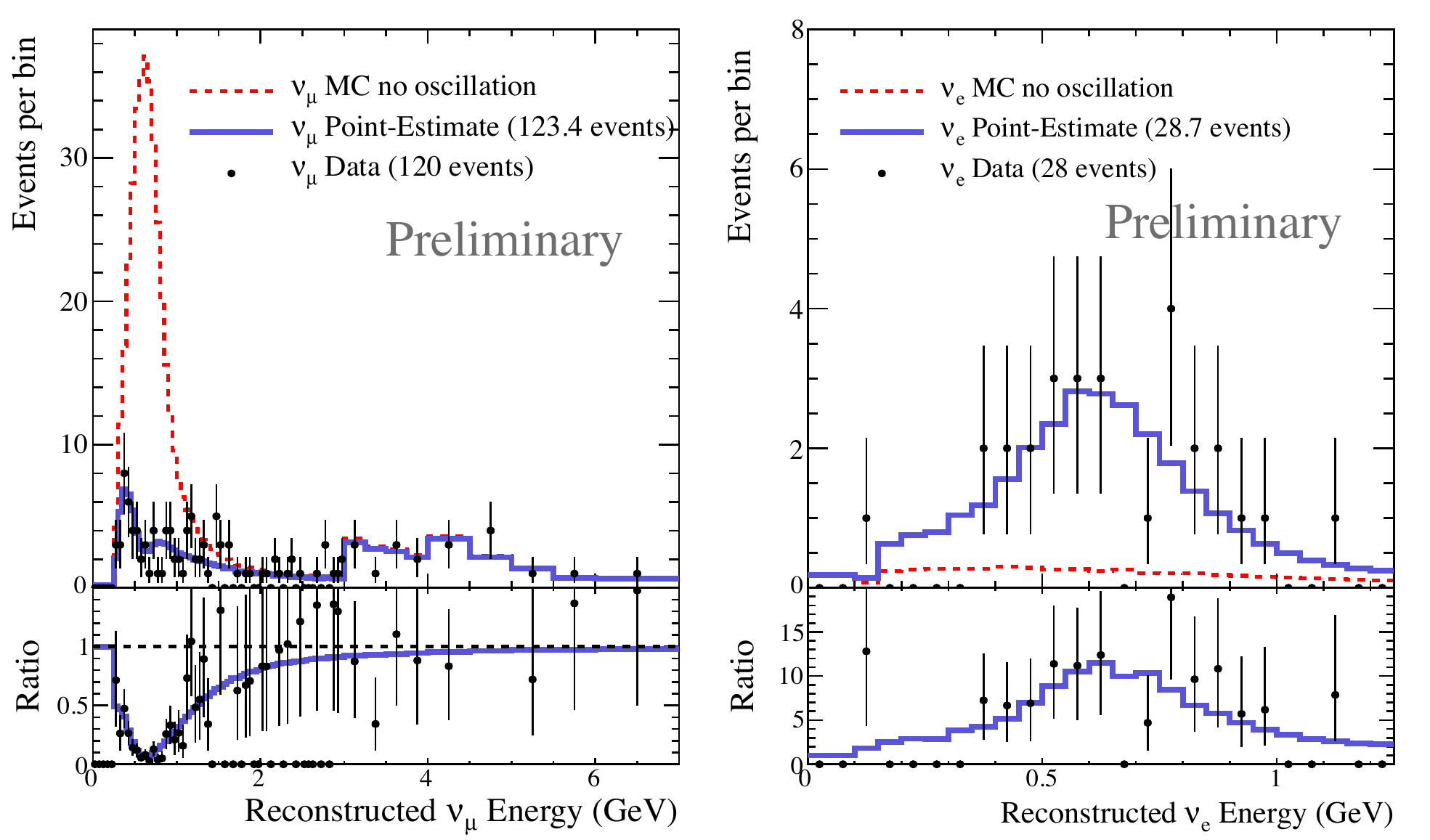}
\vspace{-0.1in}
\caption[Data and Point-Estimate Event Rates for SK Samples in $E_{rec}$]{
  The point-estimates (using T2K data only) and non-oscillated MC for the SK \num
  (left) and \nue (right) event rates compared to data. The total
  integrated numbers of events are shown in () brackets. 
  The ratios are the point-estimate over the no oscillation spectrum. 
}
\label{fig:rdf_bfs}

\end{figure}

The likelihood function $\mathcal{L}$ for the T2K oscillation analysis
is given by 
\begin{equation}
\label{eq:like_bay}
\begin{split}
  \mathcal{L}(\mathbf{o},\mathbf{b},\mathbf{x},\mathbf{d_{ND}},\mathbf{d_{SK}}|\mathbf{M}_{ND280},\mathbf{M}_{T2K-SK}) = \\ 
  P(\mathbf{M}_{ND280}|\mathbf{b},\mathbf{x},\mathbf{d_{ND}}) \times P(\mathbf{M}_{T2K-SK}|\mathbf{o},\mathbf{b},\mathbf{x},\mathbf{d_{SK}})\\
   \times \pi(\mathbf{o}) \times \pi(\mathbf{b}) \times \pi(\mathbf{x}) \times \pi(\mathbf{d_{ND}}) \times \pi(\mathbf{d_{SK}}),
\end{split}
\end{equation}
where $\mathbf{o}$, $\mathbf{b}$, $\mathbf{x}$, $\mathbf{d_{ND}}$ and
  $\mathbf{d_{SK}}$ are the neutrino oscillation, flux, cross 
section, ND280 detector and SK detector model parameters, respectively. The prior
constraints, $\pi$, for the systematic (nuisance) parameters are
described in Section~\ref{sec:t2k} and typically assume multi-dimensional Gaussians,
including correlations. For the oscillation 
parameters $|\Delta m^2_{32}|$, $\sin^2\theta_{23}$,
$\sin^2\theta_{13}$ and $\delta_{CP}$, a flat prior is
assumed. Prior constraints from solar and reactor neutrino
experiments~\cite{bib:pdg2013} are also used, in particular
$\sin^22\theta_{13} = 0.095 \pm 0.01$, which are assumed to be 
Gaussian. The prior probability for the MH (sign of $\Delta
m^2_{32}$) is 0.5 for NH and IH, except where otherwise noted. The
conditional probabilities, $P$, are assumed to be 
Poissonian and depend on the data and model predictions for each
kinematic bin in each sample as follows. The ND280
samples, $\mathbf{M}_{ND280}$, are binned in two kinematic variables (2D),
muon momentum and angle relative to the beam, as shown in 
Figures~\ref{fig:pfmom}~and~\ref{fig:pfang} for projections onto each
variable. The SK samples, $\mathbf{M}_{T2K-SK}$, are binned in reconstructed
neutrino energy assuming a CCQE interaction as shown in
Figure~\ref{fig:rdf_bfs}. The data set 
used corresponds to $0.657 \times 10^{21}$ protons on target (POT). Oscillation and
systematic parameters are varied simultaneously across all samples
when calculating the likelihood.

A Bayesian analysis is performed using an MCMC to sample the likelihood in
Equation~\ref{eq:like_bay} and marginalize over the nuisance
parameters, producing posterior probability distributions for the 
oscillation parameters of interest. This differs from the previous
T2K \nue appearance~\cite{bib:t2k_nueapprun1-4} and \num
disappearance~\cite{bib:t2k_numurun1-4} analyses that use 
frequentist-based methods such as minimization, profile-likelihoods 
and Feldman-Cousins contour construction. Another difference is the
simultaneous inclusion of ND280 and SK data in
Equation~\ref{eq:like_bay}, where the previous analyses first fit the
ND280 data only, then propagate the resulting parameter constraints to the SK
fit under a Gaussian assumption. This assumption is not necessary in a
simultaneous near and far analysis where any non-Gaussianities are
preserved in the posterior probability
distributions. There are some non-Gaussian systematics due to,
for example, nuclear model uncertainties, however the difference
between methods in the oscillation parameter estimates is currently
negligible.

A posterior probability distribution for some parameter(s) of interest
$\mathbf{p}$ is constructed by marginalizing over the nuisance
parameters $\mathbf{f}$, i.e. by integrating
Equation~\ref{eq:like_bay}: 
\begin{equation}
\label{eq:marg_like}
\mathcal{L}_{M}(\mathbf{p}|\mathbf{M}) = \int P(\mathbf{M}|\mathbf{p},\mathbf{f}) \times \pi(\mathbf{f})d\mathbf{f}.
\end{equation}
In practice, this is projecting all the steps of the MCMC onto the
parameter(s) of interest, which can be model parameters or derived
quantities. A \textit{point-estimate}, in contrast to the 
\textit{best-fit} from a minimization analysis, of $\mathbf{p}$ is determined
from the mode (most probable value) of the
$\mathcal{L}_{M}(\mathbf{p}|\mathbf{M})$
distribution. Figures~\ref{fig:pfmom}~to~\ref{fig:rdf_bfs} show the 
point-estimates for each kinematic bin in each sample (derived
quantities) compared to the data and the default predictions prior to
inclusion of the data.

\begin{figure}[htbp]
\centering
\includegraphics[width=3in]{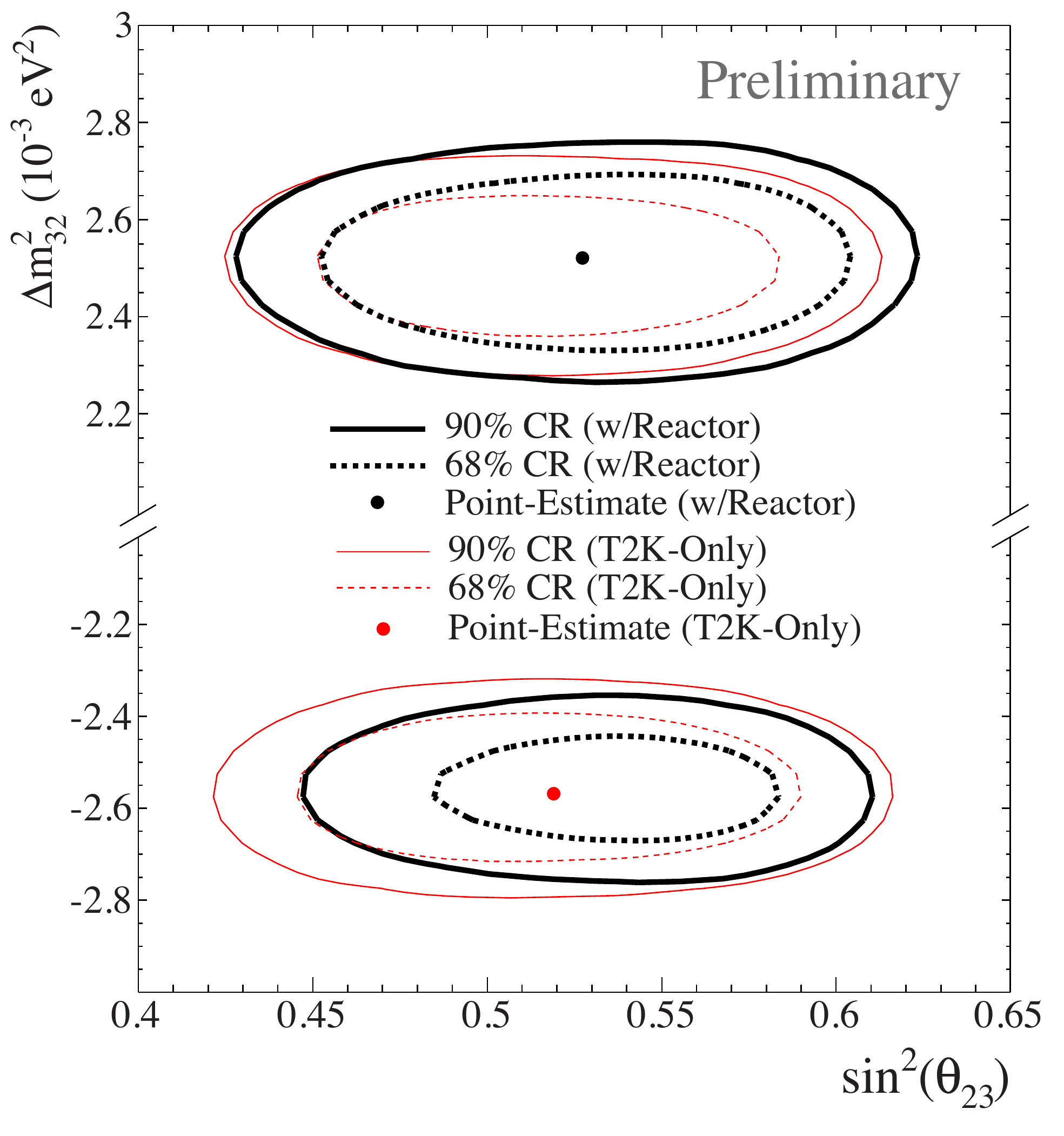}~
\includegraphics[width=3in]{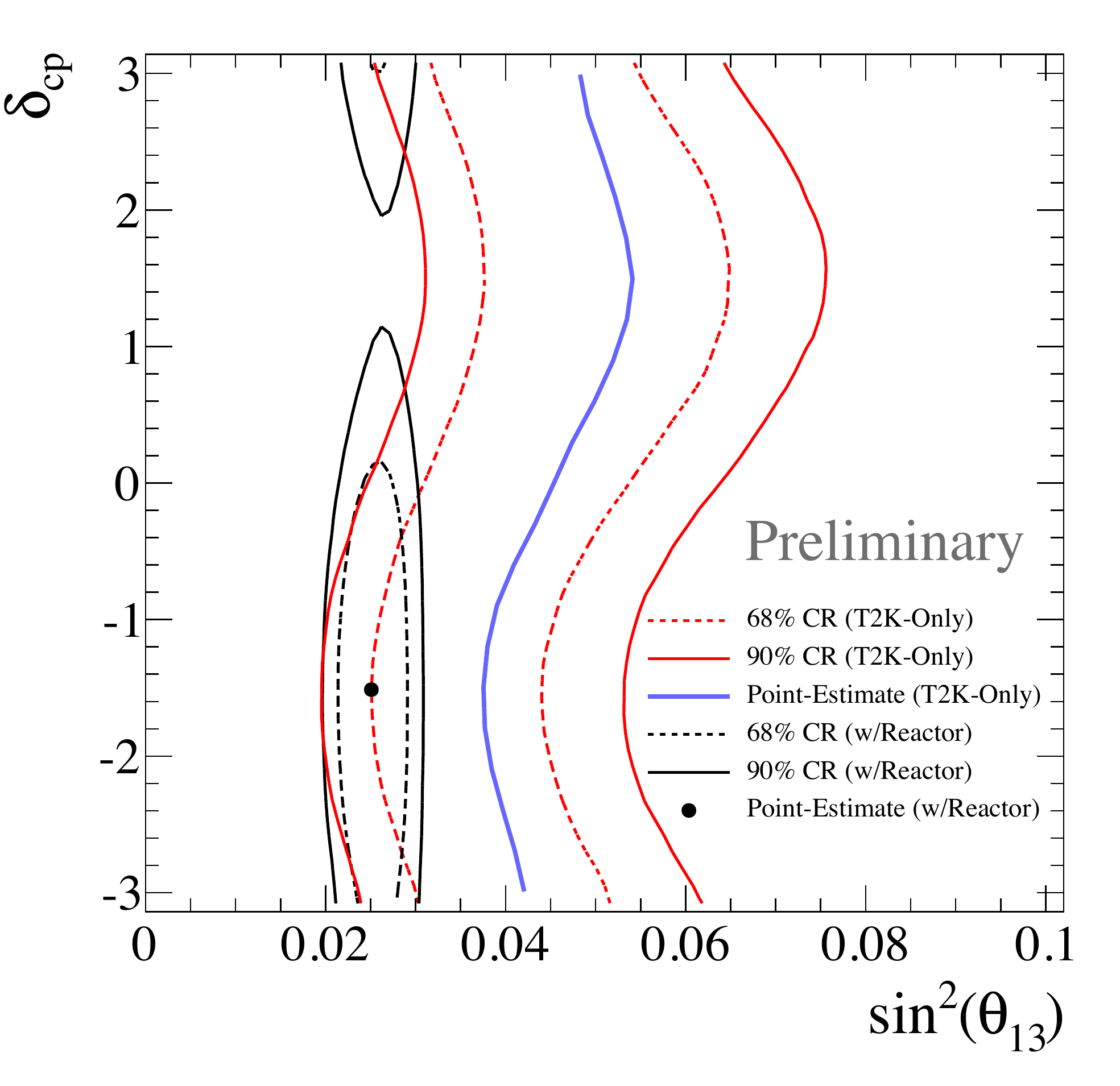}
\vspace{-0.35cm}
\caption[2D Credible Regions and Point-Estimate for Oscillation Parameters]{
  The CRs in $\Delta m^2_{32}$-$\sin^2\theta_{23}$
  (left) and $\delta_{CP}$-$\sin^2\theta_{13}$ (right) parameter
  spaces for T2K-only and reactor constrained analyses. The
  point-estimate for T2K-only in the left panel assumes
  $\delta_{CP}=0$. Note the $\Delta m^2_{32}$ axis includes the MH and
  so is not marginalized over the MH. Thus, the CRs also represent the
  probability of each MH, similar to the CL contours based on a global
  minimum in Figure~\ref{fig:numu_comparison}. 
}
\label{fig:2d_cr}
\end{figure}

A 2D $X\%$ \textit{credible region} (CR) is defined as the region
wherein the true parameter values lie with $X\%$ probability, marginalized over all 
the other parameters. It is calculated by projecting the
MCMC steps onto two parameters of interest $\mathbf{p}$, as in Equation~\ref{eq:marg_like}, then
finding the region such that
\begin{equation}
\label{eq:cr}
  X = \iint\limits_{\mathrm{CR}} \, \mathcal{L}_{M}(\mathbf{p}|\mathbf{M}) d\mathbf{p},
\end{equation}
where $\mathcal{L}_{M}$ is normalized to unity. The 68\% and 90\% CRs
are shown for the $\Delta m^2_{32}$-$\sin^2\theta_{23}$ and
$\delta_{CP}$-$\sin^2\theta_{13}$ parameter spaces in
Figure~\ref{fig:2d_cr} for the two cases of with and without the
reactor constraint. The point-estimates in 4D ($\sin^2\theta_{13}$,
$\sin^2\theta_{23}$, $\Delta m^2_{32}$, $\delta_{CP}$) are also shown
and summarized in Table~\ref{tab:summarytab}, where the 1D credible
intervals (CIs) are evaluated as in Equation~\ref{eq:cr} except for
one parameter only. The point-estimate line
for $\delta_{CP}$-$\sin^2\theta_{13}$ without the reactor constraint
is determined by scanning $\delta_{CP}$, finding the mode in 3D then
interpolating those points.

\begin{table}[htbp]
\caption[Point-Estimate and 68\% 1D Credible Intervals for the Oscillation Parameters]{
  The 4D point-estimates and 1D 68\% CIs of each oscillation
  parameter with and without the reactor constraint. 
  The CIs are calculated assuming the preferred MH, i.e. with prior
  probability $\pi(\mathrm{IH})=1$ or $\pi(\mathrm{NH})=1$. The
  probabilities for the preferred MHs are shown in
  \tab{tab:dm_model_comp}. The exclusion region for $\delta_{CP}$ is
  shown in Figure~\ref{fig:dcpoctant}.  
}
\vspace{0.2cm}
\centering
    \begin{tabular}{|c|c|c|c|c|c|}
\hline 
  ~ & Pref. MH & $|\Delta m^{2}_{32}|$ [$\times 10^{-3}$ eV$^2$] & $\sin^{2}\theta_{23}$ & $\sin^{2}\theta_{13}$ & $\delta_{CP}$ \\ \hline
    T2K-only & IH & $2.57\pm0.11$ & 0.520$^{+0.045}_{-0.050}$  & 0.0454$^{+ 0.011}_{-0.014}$ & 0 (fixed)\\ \hline
     with reactor & NH& $2.51\pm0.11 $& 0.528$^{+0.055}_{-0.038}$ & 0.0250$\pm 0.0026$ & -1.601\\
     \hline
    \end{tabular}
\label{tab:summarytab}
\end{table}

\begin{figure}
\centerline{\includegraphics[width=6in]{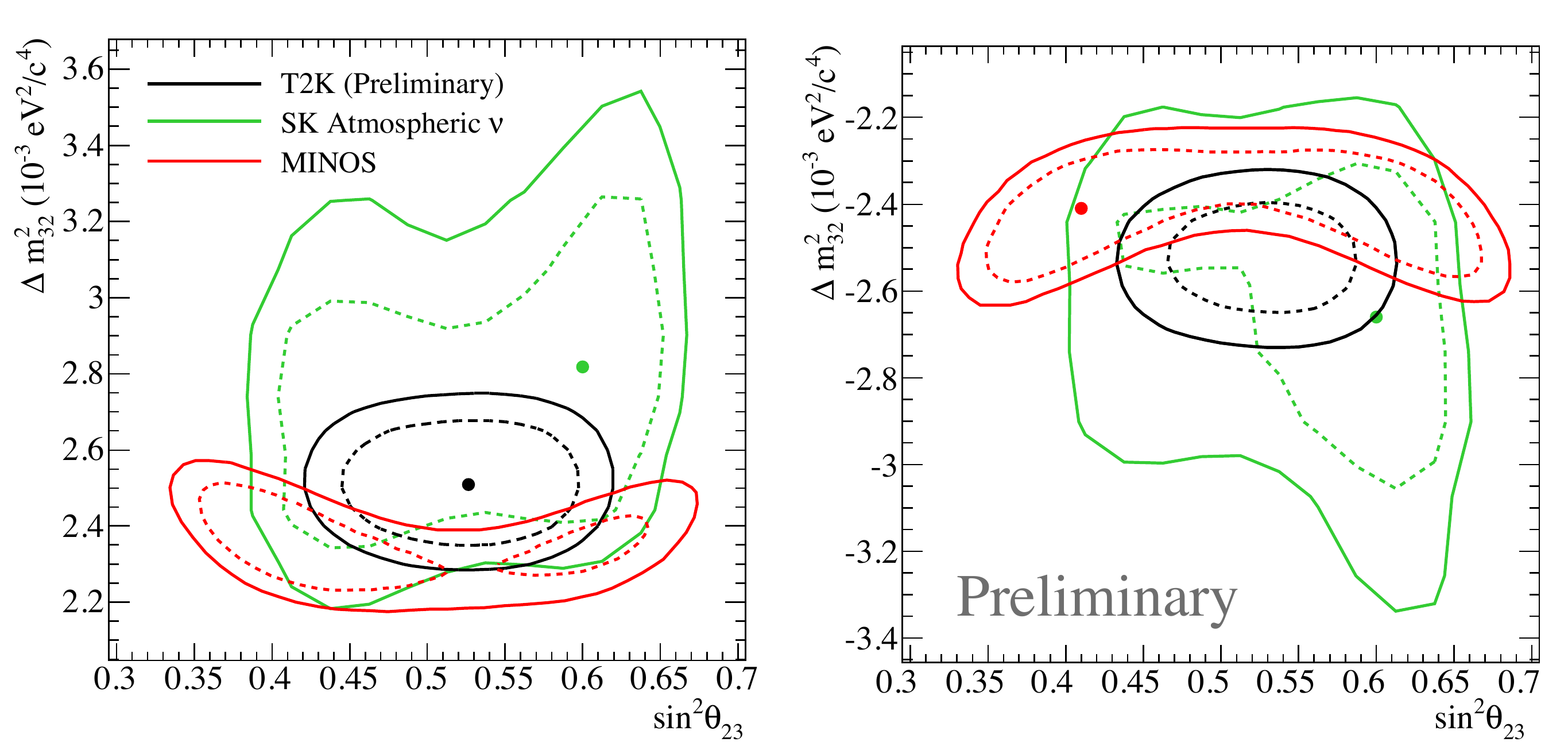}}
\vspace{-0.25cm}
\caption[]{
  The T2K best-fit point and constant-$\Delta\chi^2$ 68\% (dashed) and
  90\% (solid) CL contours in $\Delta m^2_{32}$-$\sin^2\theta_{23}$
  for NH (left) and IH (right) compared to those
  from the SK atmospheric~\cite{bib:sk_atm2013} and
  MINOS~\cite{bib:minos_finalcombined} analyses. Note the T2K and
  MINOS analyses assume a global minimum across both MHs,
  while the SK analysis presents two independent fits. 
}
\label{fig:numu_comparison}
\end{figure}

A frequentist-based analysis was also produced. The best-fits
for the oscillation parameters after minimizing over all parameters is
shown in Table~\ref{tab:summarytabvalor} for the NH and IH
assumptions. The errors are based on the 1D constant-$\Delta \chi^2$
profile for each parameter. A comparison of the $\Delta
m^2_{32}$-$\sin^2\theta_{23}$ best-fits and 68\% and 90\% confidence
level (CL) contours between the T2K, SK
atmospheric~\cite{bib:sk_atm2013} and
MINOS~\cite{bib:minos_finalcombined} analyses is shown in
Figure~\ref{fig:numu_comparison}.

\begin{table}[htbp]
\caption{
  Best-fits and 1D constant-$\Delta \chi^2$ 68\% confidence intervals
  (errors) for the oscillation parameters assuming each MH with and
  without the reactor constraint. $\Delta m^2_{32}$~($\Delta
  m^2_{13}$) is used for the NH (IH) assumption. The errors are not
  shown for $\delta_{CP}$ in the T2K-only case since there is no
  strong constraint. The errors for the other parameters in the
  reactor-constrained case are not yet calculated and will be shown in
  a future publication, while the exclusion region for $\delta_{CP}$
  is shown in Figure~\ref{fig:dcpoctant}. 
}
\vspace{0.2cm}
\centering
    \begin{tabular}{|c|c|c|c|c|c|}
\hline 
   & MH & $|\Delta m^{2}_{32,13}|$ [$\times 10^{-3}$ eV$^2$] & $\sin^{2}\theta_{23}$ & $\sin^{2}\theta_{13}$ & $\delta_{CP}$ (rad)\\ \hline
    \multirow{2}{*}{T2K-only} & NH & $2.51^{+0.11}_{-0.12}$ & 0.524$^{+0.057}_{-0.059}$  & 0.0422$^{+ 0.0128}_{-0.0212}$ & 1.9 \\ 
                              & IH & $2.49\pm0.12$ & 0.523$^{+0.073}_{-0.065}$  & 0.0491$^{+ 0.0149}_{-0.0211}$ & 1.0 \\ \hline
    
    \multirow{2}{*}{Reactor-constrained} & NH &  $2.51$ & 0.527 & 0.0248 & -1.55 \\
     & IH & $2.48$ & 0.533 & 0.0252 & -1.56 \\
   
    \hline
    \end{tabular}
\label{tab:summarytabvalor}
\end{table}

\begin{figure}[htbp] 
   \centering

   \centerline{\includegraphics[width=3.2in]{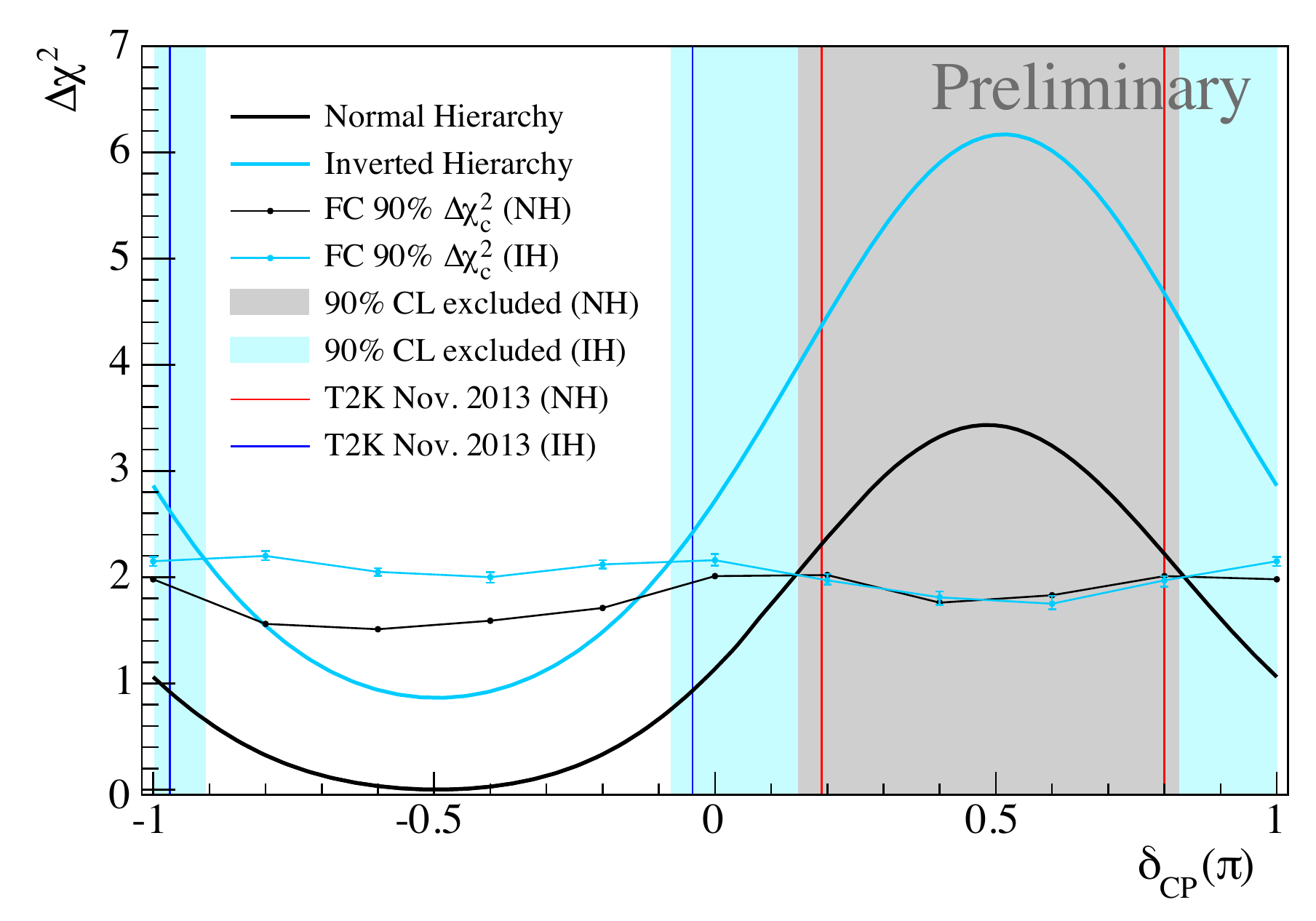}~\includegraphics[width=3.2in]{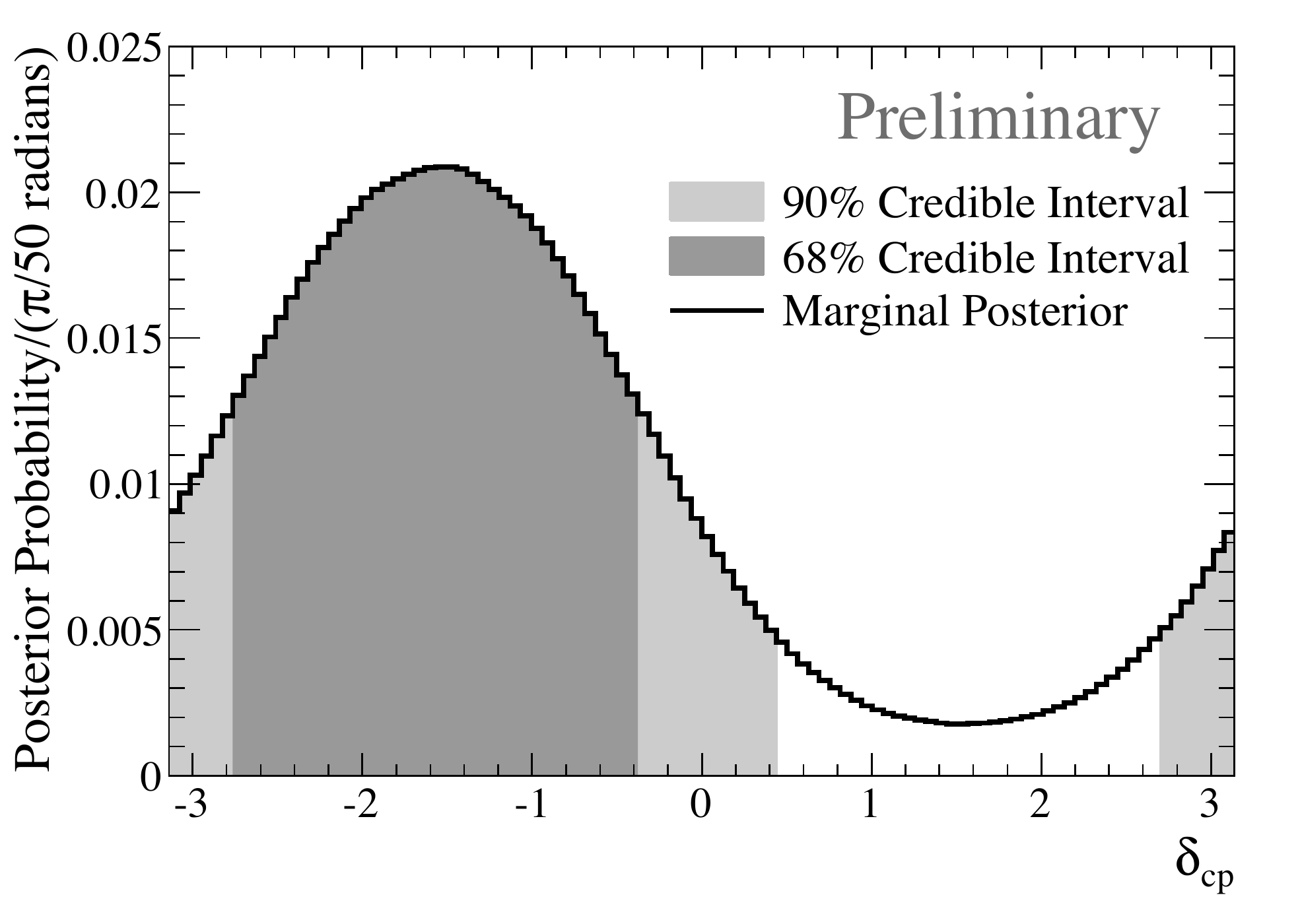}}
\vspace{-0.4cm}
   \caption[$\delta_{CP}$ Posterior Distribution with Reactor Constraint]{
     Left: The $\Delta\chi^2$ profile for $\delta_{cp}$, showing the
     90\% CL regions based on the FC $\Delta \chi^2_c$
     for NH and IH. The The 90\% CL inclusion is $[-1.18, 0.15]\pi$
     for NH and $[-0.91, -0.08]\pi$ for IH.  The vertical lines show
     the previous \nue appearance
     result~\cite{bib:t2k_nueapprun1-4}. Right: The posterior
     probability and 68\% and 90\% CIs for $\delta_{CP}$, marginalized
     over all the other parameters including the MH with priors
     $\pi(\mathrm{NH})=\pi(\mathrm{IH})=0.5$. The 90\% CI is $[-1.11, 0.38]\pi$.   
   }
   \label{fig:dcpoctant}
 \end{figure}

Allowed intervals for $\delta_{CP}$ with the reactor constraint are shown in
Figure~\ref{fig:dcpoctant} for the frequentist-based analysis
including a Feldman-Cousins (FC) critical $\Delta \chi^2$ correction
($\Delta \chi^2_c$) and the Bayesian analysis using the posterior
probability and CIs. Referring to Equation~\ref{eq:3flav_app_prob},
$\delta_{CP} \approx -\pi/2$ is preferred since the T2K data alone
prefers a larger $\theta_{13}$ compared to the reactor data.

The Bayesian analysis can also make interesting, though not
yet significant, statements about the MH and $\theta_{23}$ 
octant. Table~\ref{tab:dm_model_comp} shows the posterior 
probabilities for each MH and $\theta_{23}$ octant combination. The
normal hierarchy and first octant is preferred when including the
reactor constraint, similarly to the $\delta_{CP}$ preference
above. The MH, $\theta_{23}$ octant and $\delta_{CP}$ result here is
in slight tension with the MINOS result~\cite{bib:minos_finalcombined}.

\begin{table}[htbp]

\caption[Posterior Probabilities for each MH and $\theta_{23}$ Octant Combination]{
  The posterior probability for each MH and $\theta_{23}$ octant
  combination, assuming prior probabilities
  $\pi(\mathrm{NH})=\pi(\mathrm{IH})=0.5$ and
  $\pi(\sin^2\theta_{23}<0.5)=\pi(\sin^2\theta_{23}>0.5)=0.5$. All other
  parameters are marginalized without (left) and with (right) the
  reactor constraint. 
}
\label{tab:dm_model_comp}
\vspace{0.2cm}
\centering
    \begin{tabular}{|c|cc|c|} 
\multicolumn{4}{c}{T2K-Only}\\
\hline
    & NH & IH & Sum\\	
    \hline
    $\sin^2\theta_{23}\leq0.5$&0.165&0.200&0.365\\
    $\sin^2\theta_{23}>0.5$&0.288&0.347&0.635\\
    \hline
    Sum & 0.453& 0.547&\\
\hline
    \end{tabular}~~~~~~~~~~~~~\begin{tabular}{|c|cc|c|} 
\multicolumn{4}{c}{Reactor-Constrained}\\
\hline 
    & NH & IH & Sum\\	
    \hline
    $\sin^2\theta_{23}\leq0.5$&0.179&0.078&0.257\\
    $\sin^2\theta_{23}>0.5$&0.505&0.238&0.743\\
    \hline
    Sum & 0.684& 0.316&\\
\hline 
    \end{tabular}
\end{table}

\vspace{-0.1in}

\section{Conclusion and Outlook}

The first T2K combined \num disappearance and \nue appearance analyses
based on $0.657 \times 10^{21}$ POT is presented. T2K is producing the
leading measurement on $\theta_{23}$ and, combined with reactor
neutrino data, non-trivial exclusion intervals in $\delta_{CP}$. The
first anti-neutrino run is scheduled this year and, with increasing
POT in the coming years, T2K will continue to lead the search for CP
violation in the lepton sector.


\end{document}